\documentclass{imsart}

\RequirePackage{amsthm,amsmath,amsfonts,amssymb}
\RequirePackage[authoryear]{natbib}
\RequirePackage{graphicx}

\startlocaldefs

\endlocaldefs

\begin{document}

\begin{frontmatter}
\title{Estimating effects within nonlinear autoregressive models: a case study on the impact of child access prevention laws on firearm mortality\thanksref{T1}}
\thankstext{T1}{Support for this research was from a grant from Arnold Ventures.}

\begin{aug}
\author[A]{\fnms{Matthew} \snm{Cefalu}},
\author[A]{\fnms{Terry} \snm{Schell}},
\author[A]{\fnms{Beth Ann} \snm{Griffin}},
\author[A]{\fnms{Rosanna} \snm{Smart}},
\and
\author[A]{\fnms{Andrew} \snm{Morral}}

\address[A]{RAND Corporation}

\end{aug}

\begin{abstract}
Autoregressive models are widely used for the analysis of time-series data, but they remain underutilized when estimating effects of interventions. This is in part due to endogeneity of the lagged outcome with any intervention of interest, which creates difficulty interpreting model coefficients. These problems are only exacerbated in nonlinear or nonadditive models that are common when studying crime, mortality, or disease. In this paper, we explore the use of negative binomial autoregressive models when estimating the effects of interventions on count data. We derive a simple approximation that facilitates direct interpretation of model parameters under any order autoregressive model. We illustrate the approach using an empirical simulation study using 36 years of state-level firearm mortality data from the United States and use the approach to estimate the effect of child access prevention laws on firearm mortality. 

\end{abstract}


\end{frontmatter}

\section{Introduction}

Autoregressive models and their extensions are powerful tools for the analysis of time-series data. These approaches have a rich history in forecasting \citep{de200625,granger2014forecasting}, econometrics \citep{lutkepohl2004applied,hamilton2020time}, and statistics \citep{box2015time}. Despite the widespread use of these methods, autoregressive models remain underutilized when estimating the effect of interventions, in part due to the difficulty interpreting the model parameters. The parameters from a standard autoregressive model are difficult to interpret because these models control for the lagged outcome, a variable which is endogenous to the intervention of interest. Because of this endogeneity, the estimated coefficient relating the intervention to the outcome represents only the fraction of the overall policy effect that is not mediated through the autoregressive effect \citep{pearl2009causality}. If the regression coefficient within an autoregressive model were to be interpreted as the intervention's effect, it would be biased relative to the true effect \citep{nickell1981biases,achen2000lagged}.  

For linear and additive autoregressive models, there is a literature on estimating the relationship of one time series to another. These approaches typically involve reparameterizing the autoregressive model to ensure that the model assumptions are met and that the model parameters represent the effect of interest, or to use vector autoregressive models \citep{engle1987co,johansen1991estimation,keele2004not,lutkepohl2004applied}. A prominent example of using reparameterization is co-integration and error correction in the autoregressive distributed lag model \citep{pesaran1998autoregressive,dufour1998exact,hassler2006autoregressive,philips2018have}. While these approaches and concepts are well developed for linear and additive autoregressive models, guidance for non-linear or non-additive models is unclear. This is a particular challenge when studying incident data, such as counts of crime, mortality or disease, where linear models often make untenable assumptions, and models with nonlinear link functions and non-normally distributed error are more appropriate and more commonly used.  

Methods for modeling autoregressive incidence data vary and include approaches that treat the counts as continuous and use traditional autoregressive models, approaches that include the lagged count into the mean function \citep{fokianos2009poisson}, approaches that model the counts as mixtures of Poisson processes \citep{mckenzie1986autoregressive}, and state-space approaches that postulate a measurement model and a transition equation \citep{brandt2000dynamic}. With such varied approaches, there is little clarity on how best to model autoregressive count data and much less on how to do so in a way that ensures model coefficients are unbiased and can be interpreted as effect estimates. 

Due to this complexity, methodological guidance for estimating intervention effects in time-series data has often been to avoid autoregressive models, and use other approaches such as difference-in-differences methods \citep{shang2018interaction,kondo2015difference,angrist2008mostly} and synthetic control methods \citep{abadie2010synthetic,ben2021synthetic,arkhangelsky2019synthetic}. However, a recent simulation study highlighted that many of these alternate approaches can suffer inflated Type-I error rates, bias, and extremely low power when estimating state-level policy effects on firearm mortality data \citep{schell2018evaluating}. The authors identified a negative binomial autoregressive model with the state-level policy effects coded as first differences as being the optimal choice based on its empirical performance relative to the other methods considered. However, the recommended approach still showed a small amount of bias, was not clearly generalizable to other outcomes, and does not have a simple extension beyond the first order autoregressive model. Subsequent work by \citet{schell2020} improved on this approach by using numerical integration to estimate marginal effects that are interpretable, but this procedure is computationally burdensome and difficult for non-technical audiences to understand. Moreover, it is unclear how higher-order autoregressive lags could be added to this model.

In this paper, we explore the use of a modification of autoregressive models for count data originally proposed in \citet{zeger1988markov} for Poisson models, and extended to generalized autoregressive moving average models by \citet{benjamin2003generalized}. We show that a simple approximation facilitates direct interpretation of the model parameters and allows for any order autoregressive coefficient, overcoming a major limitation of using autoregressive models when analyzing the effects of state or local policies on crime data. We perform an empirical simulation study to verify the properties of the approach using 36 years of state-level firearm mortality data from the United States. The simulation was designed to evaluate these methods in a complex nonlinear context: we use a negative binomial model, estimate effects that change over time, and investigate the approach using both first- and second-order autoregressive models. Finally, we present a case study demonstrating that the model recovers substantively equivalent estimates for one state gun law as found in \citet{schell2020}. 

\section{Methods} \label{sec:methods}
\subsection{Linear Autoregressive Models}

We begin by defining effects of interest as the effects of a policy at specific times after implementation. While we focus our presentation on binary policy indicators, the effects and models discussed in this paper are appropriate for other variables of interest, e.g., continuous variables. Let $y_{it}$ be an outcome for observation $i$ at time $t$, with $i=1,\dots,N$ and $t=0,\dots,T$. Let $A_{it}$ denote an indicator of a policy of interest being active for observation $i$ at time $t$. Denote the vector of $A_{it}$ for time $0$ to $t$ as $\mathbf{A}_{it}$. We begin by defining an effect of interest at time $t$ given :
\begin{align*}
   \Delta_t(\mathbf{a}) = E[ y_{it} | \mathbf{A}_{it} = \mathbf{a} ] - E[ y_{it} | \mathbf{A}_{it} = \mathbf{0} ] ,
\end{align*}

\noindent where $\mathbf{a}$ is a sequence of interest of $A_{it}$. For example, $\mathbf{a}$ could denote the implementation of a gun law at $t=1$ such that $\mathbf{a}=(0 , 1 , \dots , 1)$. In this case, $\Delta_t(\mathbf{a})$ would represent a comparison of the expected outcome at time $t$ with the policy implemented at time $1$ to the expected outcome at time $t$ had the policy not been implemented up through time $t$.

Note that the effects defined here are a deviation from the typical short-run or long-run interpretation of autoregressive model coefficients. Instead, these effects are defined at specific time points given a hypothetical implementation of a policy, which provides actionable information about the effect of a policy in the short- and mid-run. These effects and the results of this paper can be generalized to allow for a comparison of two arbitrary policy sequences by comparing $E[ y_{it} | \mathbf{A}_{it} = \mathbf{a} ]$ to $E[ y_{it} | \mathbf{A}_{it} = \mathbf{a}' ]$ for two sequences of interest of $A_{it}$, $\mathbf{a}$ and $\mathbf{a}'$.


As an illustrative example, consider the linear autoregressive model of order $1$, such that:
\begin{align*}
    y_{it} &= \alpha + \delta y_{i,t-1} + \theta A_{it} + \epsilon_{it},
\end{align*}

\noindent where $\epsilon_{it}$ are independent with mean 0. Under this model, $\Delta_t(\mathbf{a})$ can be shown to be:
\begin{align*}
    \Delta_t(\mathbf{a}) = \theta \sum_{k=0}^t \delta^k a_{t-k}.
\end{align*}

This highlights the main limitation of the linear autoregressive model for estimating the effect of a policy -- the effect is a function of the autoregressive coefficient, the regression coefficient on the policy, and the policy sequence of interest. We consider a special case of \citet{benjamin2003generalized} that avoids these limitations and has been discussed elsewhere \citep{zeger1988markov}, which we will refer to as the debiased autoregessive model (DAM):

\begin{align}
    y_{it} &= \alpha + \delta (y_{i,t-1} - \theta A_{i,t-1} ) + \theta A_{it} + \epsilon_{it}. \label{eqn:linear}
\end{align}

\begin{figure}
    \centering
    \includegraphics[width=.49\textwidth]{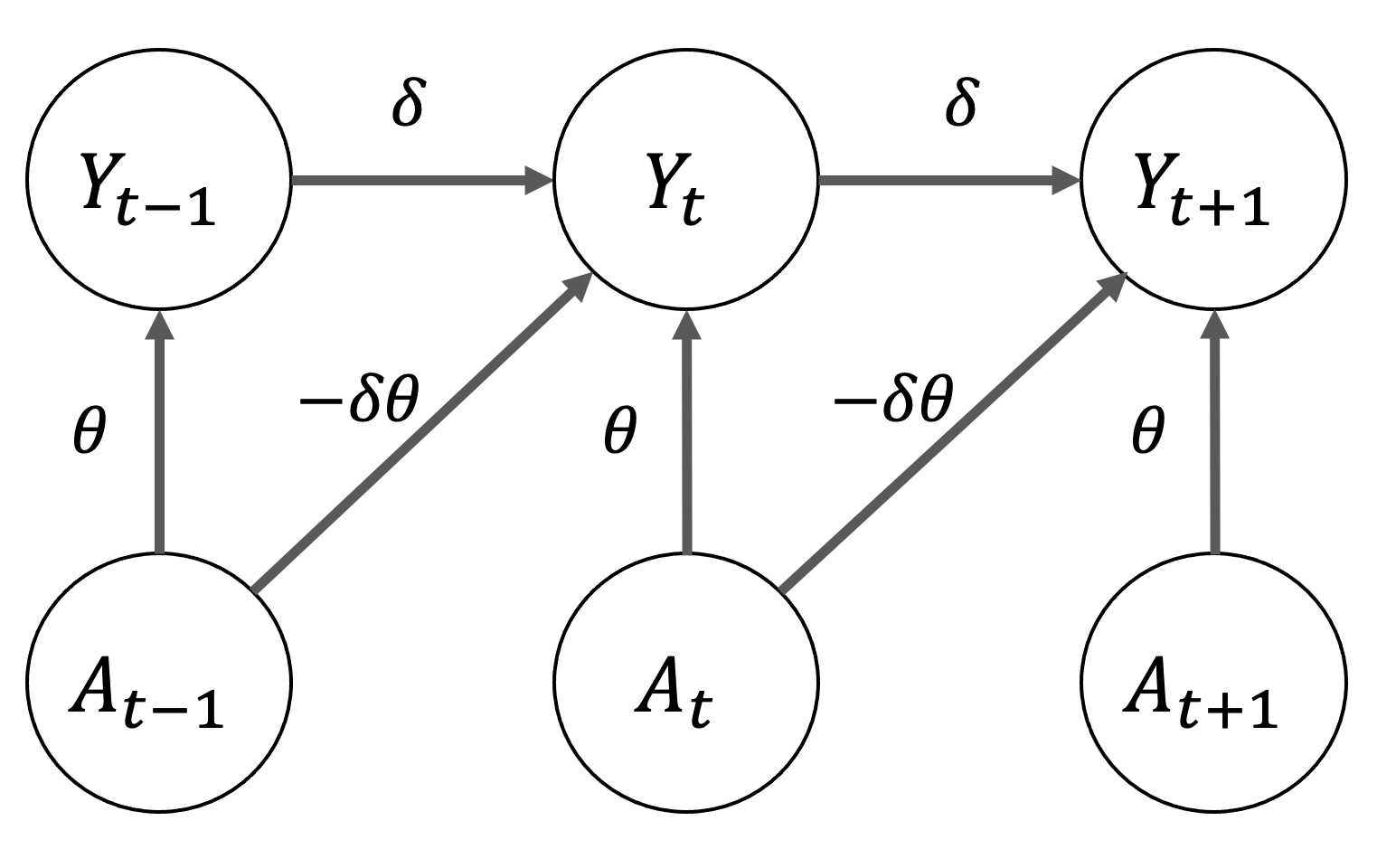} 
    \caption{Graphical representation of a linear debiased autoregressive model.}
    \label{fig:path}
\end{figure}

Figure \ref{fig:path} provides a graphical representation of this model. Note that the indirect path from $A_{t-1}$ to $Y_{t}$ through $Y_{t-1}$ is cancelled out by the direct path from $A_{t-1}$ to $Y_{t}$. This removes the accumulation of the policy effect over time through these indirect paths, leads to a model coefficient that is directly interpretable, and motivates the terminology ``debiased'' autoregressive model. Specifically, under the DAM specification it can be shown that 

\begin{align*}
    \Delta_t(\mathbf{a}) = \theta a_{t}.
\end{align*}

\noindent Here, we point out that the effect of interest is defined only by the regression coefficient and the value of the policy at time $t$. In this way, the model coefficients provide a directly interpretable effect as the total effect of policy at time $t$.  

The DAM can be generalized to allow for higher order autoregressive terms, an arbitrary function of the policy vector, and inclusion of covariates. Let $\mathbf{A}_{it} = (A_{i0},A_{i1},\dots , A_{it}) $ be the history of the policy up to time $t$, and let $\mathbf{X}_{it}=(\mathbf{X}^{(1)}_{it},\mathbf{X}^{(2)}_{it})$ be similarly defined covariate histories up to time $t$. Let $f_t(\cdot)$, $g_t(\cdot)$, and $h_t(\cdot)$ be arbitrary scalar functions, and let $m_t(\mathbf{A}_{i,t},\mathbf{X}^{(1)}_{i,t}) = f_{t}(\mathbf{A}_{i,t}) + g_{t}(\mathbf{X}^{(1)}_{i,t})$. The DAM of autoregressive order $p$, DAM($p$), is given by:

\begin{align}
    y_{it} &= \alpha + \sum_{k=1}^{p} \delta_k \left\{y_{i,t-k} - m_{t-k}(\mathbf{A}_{i,t-k},\mathbf{X}^{(1)}_{i,t-k}) \right\} \label{eqn:linearDAM} \\ 
    & ~~~~ + m_t(\mathbf{A}_{it},\mathbf{X}^{(1)}_{it}) + h_t(\mathbf{X}^{(2)}_{it}) + \epsilon_{it}. \nonumber
\end{align}

\noindent Under this specification, the effect of interest can be shown to be $\Delta_t(\mathbf{a})=f_t(\mathbf{a})$. The effect of interest is directly defined by the functions specified in the regression model and is free of all autoregressive terms. We allow the functions $f_t(\cdot)$, $g_t(\cdot)$, and  $h_t(\cdot)$ to be different at each time $t$, but in practice, most specifications will parameterize the functions to pool information across time points. One example is a distributed lag formulation of order $b$ for $f_t(\cdot)$, i.e., $f_t(\mathbf{A}_{it}) = \sum_{j=0}^{b} \theta_j A_{t-j}$. 

We partition the covariates into two sets, $\mathbf{X}^{(1)}_{it}$ and $\mathbf{X}^{(2)}_{it}$, to allow for additional flexibility in the specification of the model. The inclusion of $h_t(\mathbf{X}^{(2)}_{it})$ is flexibility beyond that allowed in \citet{benjamin2003generalized}. If $\mathbf{X}^{(2)}$ is empty, then the functions of the covariates specified in $g_{t}(\cdot)$ are interpretable in a manner similar to $f_t(\cdot)$. If $\mathbf{X}^{(1)}$ is empty, then the functions of the covariates specified in $h_{t}(\cdot)$ have properties similar to those of standard autoregressive models, e.g., the effect of the covariate is allowed to accumulate over time. 


The proposed model in (\ref{eqn:linearDAM}) includes some of the most common models for effect estimation in longitudinal data as special cases: first-differences models and two-way fixed effect models. In particular, a first-differences model is a special case of (\ref{eqn:linearDAM}) when the autoregressive coefficient is set to 1. By allowing for the estimation of the autoregressive coefficient, (\ref{eqn:linearDAM}) is appropriate in a wider range of settings where fixing the autoregressive coefficient to unity results in a violation of the model assumptions and poor effect estimates.  

Similarly, the standard two-way fixed effect model falls out as a special case whenever time and unit fixed effects are added to (\ref{eqn:linearDAM}). If the assumptions of the standard two-way fixed effect are met, i.e., errors are independent conditionally on the fixed effects, the autoregressive term will be zero. Thus, when the standard two-way fixed effects model assumptions are met in the sample, (\ref{eqn:linearDAM}) with time and unit fixed effects reduces to the two-way fixed effect model. However, allowing the model to estimate the autoregressive term should allow the model to be more appropriate in a wider range of settings than the standard two-way fixed effects model.

\subsection{Extensions to Negative Binomial Models}


To facilitate the use of a negative binomial regression model with a log link function, the previously defined effect is modified to be a rate ratio instead of rate difference. That is, 

\begin{align}
   \Delta_t(\mathbf{a}) = \frac{E[ y_{it} | \mathbf{A}_{it} = \mathbf{a} ]}{E[ y_{it} | \mathbf{A}_{it} = \mathbf{0} ]}. \label{eqn:rr}
\end{align}


Motivated by the previous results, we describe two specifications of negative binomial models with similar form for the linear predictor of the expectation from (\ref{eqn:linear}) and (\ref{eqn:linearDAM}). First, an autoregressive negative binomial of order $p$ is given by:
\begin{align}
     y_{it} &\sim NegBin(\mu_{it},\phi) \nonumber \\
     \log(\mu_{it}) &=  \alpha + \sum_{k=1}^{p} \delta_k  \log(y_{i,t-k} / N_{i,t-k}) + m_t(\mathbf{A}_{it},\mathbf{X}^{(1)}_{it}) + h_t(\mathbf{X}^{(2)}_{it}) + \log(N_{it})   \label{eqn:nb-adl}
\end{align}

Under this model, a simple closed form for the effects of interest in (\ref{eqn:rr}) is unlikely to exist except in special cases. Numerical integration approaches can be used to estimate the expectations in (\ref{eqn:rr}); see \citet{schell2020} for an example application. Second, a negative binomial DAM of order $p$, NB-DAM($p$), is given by:

\begin{align}
     \log(\mu_{it}) &=  \alpha + \sum_{k=1}^{p} \delta_k \left( \log(y_{i,t-k} / N_{i,t-k}) - m_{t-k}(\mathbf{A}_{i,t-k},\mathbf{X}^{(1)}_{i,t-k}) \right)      \label{eqn:nb-dam} \\
     & ~~~ + m_t(\mathbf{A}_{it},\mathbf{X}^{(1)}_{it}) + h_t(\mathbf{X}^{(2)}_{it}) + \log(N_{it}). \nonumber
\end{align}

A technical discussion and the properties of this model when $\mathbf{X}^{(2)}$ is empty can be found in \citet{benjamin2003generalized}. Unlike the DAM($p$), the NB-DAM($p$) does not have a closed form for the effect of interest. To avoid computation problems associated with numerical integration and to facilitate both interpretation and inference, we provide an approximation to the effect of interest as:

\begin{align} 
   \tilde{\Delta}_t(\mathbf{a}) = exp\left( f_t(\mathbf{a}) \right) \label{eqn:est_effect} 
\end{align}

This approximation can be derived by replacing $\log(y_{i,t-k})$ with its corresponding $\log(\mu_{i,t-k})$ inside of (\ref{eqn:nb-dam}). Such an approximation is not exact due to the nonlinearities in the model, but we will explore its properties using a simulation study in Section \ref{sec:sim}. 

%


The structure of the $\text{NB-DAM}(p)$ given in (\ref{eqn:nb-dam}) provides great flexibility in the specification of the policy effect on outcomes. As previously noted, one specification that maintains flexibility is a distributed lag structure, i.e., $f_t(\mathbf{A}_{it}) = \sum_{j=0}^{b} \theta_j A_{t-j}$. However, it may be of interest to constrain the structure in a parsimonious fashion to further improve interpretability and boost power. With this in mind, we propose a simple specification of $f_t(\cdot)$ that allow for the effect of a policy to be a combination of an instantaneous effect and an effect that phases in linearly (on the log rate) over a pre-specified time period. For illustration, we will assume that the policy becomes active at the beginning of the period covered by $t=0$ such that the linear phase-in effect accumulates during the period represented by $t=0$, and we assume the period of the linear phase in is $b$. Then, 

  \begin{equation}
    f_t(\mathbf{A}_{it}) = 
    \begin{cases}
      A_{i0} \left( \beta_0 + \frac{\beta_1}{2b} \right), & \text{if } t=0 \\
      f_{t-1}(\mathbf{A}_{i,t-1}) +  \frac{A_{it}\beta_1}{b}, & \text{if } 1 \leq t < b \\
      f_{t-1}(\mathbf{A}_{i,t-1}) + \frac{A_{it}\beta_1}{2b} & \text{if } t = b \\
      f_{t-1}(\mathbf{A}_{i,t-1}) & \text{if } t > b
    \end{cases} \label{eqn:approx}
  \end{equation}

Under this specification, the approximation to the total effect of a policy at year $b$ is given by:

\begin{align*}
   \tilde{\Delta}_b(\mathbf{1}) = exp\left( \beta_0 + \beta_1 \right).
\end{align*}

\noindent The Appendix provides a more nuanced discussion of these effects, including an extension that account for the fact that a law may be in effect for a fraction of the time period associated with observations at time $t$ (e.g., laws are enacted at various time points throughout the year).

Bayesian estimation of this model is straightforward and allows for the specification of priors:

\begin{align*}
    \alpha & \sim N(a_0,a_1) \\
    \delta_k & \sim U(0,1) \text{  for } k = 1, \dots , p \\
    \beta_j & \sim N(c_j,d_j) \text{  for } j = 0, 1
\end{align*}

\noindent where $a_0$, $a_1$, $c_j$, and $d_j$ are hyperparmeters controlling the level of information in the prior distributions. The choice of a uniform distribution from 0 to 1 for the autoregressive coefficients rules out values of these coefficients that would lead to models that are outside of our expectations, e.g., a value of an autoregressive coefficient that is less than 0 or greater than 1. Implications of autoregressive coefficients outside of this range can be found in \citet{benjamin2003generalized}. The choice of hyperparameters for the $\beta_j$'s can be set such that the prior distributions are noninformative, or they can be chosen to place the prior distribution in a region based on expert knowledge.

\section{Demonstration of model performance}
We demonstrate the performance of the NB-DAM in two ways. First, we simulate the effects of hypothetical laws, comparing law effect estimates recovered using the proposed model to estimates from other candidate models. Next, using genuine state law data, we demonstrate through a case study that the NB-DAM recovers estimates of the effects of child access prevention laws (CAP laws) on total firearm deaths that are substantively equivalent to using (\ref{eqn:nb-adl}) with numerical integration \citep{schell2020}.  

\subsection{Data}
Our simulation and our model of CAP law effects use longitudinal data from 50 states for the years 1980–2016. The primary outcome for these estimates are state-level counts of annual firearm deaths from the Vital Statistics System, which contains information on coroners’ cause of death determinations for over 99\% of all deaths in the U.S \citep{nchs_vital}. 

Following procedures described in \citet{schell2020}, our model of the effects of state laws on firearms deaths controls for 34 state-level time-varying demographic, economic, crime, law enforcement, and gun ownership characteristics. These include characteristics known to be associated with firearm deaths or which are are commonly used when analyzing state-level differences in health or crime. These variables are all taken from federal government sources and constitute descriptive statistics for each state for each year in the studied period. Because many of these  covariates are collinear, the model includes only the first 17 principal components extracted from the larger set of 34 covariates, which together accounted for 95\% of the variance in the full matrix of covariates. The model also includes year fixed effects.

Additional information about each variable and our dimension-reduction techniques can be found in \citet{schell2018evaluating}. 

\subsection{Simulation Study} \label{sec:sim}

This simulation study focuses on variations of negative binomial autoregressive models based on the findings of \citet{schell2018evaluating}, who recommended these models over other common approaches such as two-way fixed effects models. The synthetic control method for staggered adoption \citep{ben2021synthetic} was also considered, but using the software available at the time of writing, the approach had large variance and always estimated a null effect. We compare four alternative approaches to estimating the effect of laws, each a negative binomial model with lagged dependent variables:

\begin{enumerate}
    \item {\itshape Effect coded}: the model in (\ref{eqn:nb-adl}) with $f_t(\cdot)$ defined by (\ref{eqn:approx}) with $b=5$
    \item {\itshape Change coded}: the model in (\ref{eqn:nb-adl}) with $f_t(\cdot)$ defined by the first differences of (\ref{eqn:approx}) with $b=5$, i.e., $f_t(\mathbf{A}_{it})-f_{t-1}(\mathbf{A}_{i,t-1})$
    \item {\itshape NB-DAM(1)}: the NB-DAM(1) with $f_t(\cdot)$ defined by (\ref{eqn:approx}) with $b=5$
    \item {\itshape NB-DAM(2)}: the NB-DAM(2) with $f_t(\cdot)$ defined by (\ref{eqn:approx}) with $b=5$
\end{enumerate}

\noindent In each model, the estimated effect is extracted using the model coefficient directly; that is, no numerical integration approaches are used. The change coded model was included as it was the preferred method of \citet{schell2018evaluating}. The 17 principle components were included as linear effects in each model, and $\mathbf{X}^{(1)}$ is left empty in the NB-DAM models.

We consider a hypothetical policy that is implemented in 15 states on randomly selected years. The data generating scheme starts with the observed time-series in each state, randomly draw 15 states that implement a hypothetical policy, randomly assign dates in which the policy takes effect, and adds or subtracts firearm deaths from the observed time-series after the laws ``implementation'' based on the effect of the policy we have assigned to it. The purpose of this simulation is to verify that the proposed model in (\ref{eqn:nb-dam}) and the approximation described in (\ref{eqn:est_effect}) provide approximately unbiased inference.

For a hypothetical policy that has an instantaneous risk ratio of $e^{\beta_0}$ and an effect with risk ratio $e^{\beta_1}$ that phases in over 5 years, the steps of the empirical simulation study are as follows:

\begin{enumerate}
    \item Randomly select 15 states that implement the hypothetical policy.
    \item Random assign implementation dates for each of the 15 randomly selected states between January 1981 and December 2007.
    \item The total effect of the hypothetical policy during the first 5 years after implementation is given by (\ref{eqn:approx}) with $b=5$. The total effect of the policy is assumed to be constant after 5 years, $e^{\beta_0+\beta_1}$.
    \item In each of the 15 states, multiply the observed number of firearm deaths by the total effect for each year after the implementation of the hypothetical policy. 
    \item Analyze these data using the methods under consideration.
    \item Repeat 1,000 times.
\end{enumerate}

\noindent This procedure is repeated for all combinations of
\begin{align*}
    e^{\beta_0} &\in (0.9,0.95,0.99,1,1.01,1.05,1.1),\text{ and} \\
    e^{\beta_1} &\in (0.9,0.95,0.99,1,1.01,1.05,1.1).
\end{align*}

Table \ref{tab:null} shows the simulation results for a hypothetical policy that has no effect ($e^{\beta_0}=e^{\beta_1}=1$). All models are approximately unbiased. The effect coded model has the lowest MSE, but suffers from inflated Type I error. Both the NB-DAM(1) and NB-DAM(2) have lower MSE than the change coded model, and both have slightly higher Type I error rates than the nominal 5\%. The NB-DAM(2) has lower MSE than the NB-DAM(1). 

Table \ref{tab:power} shows the simulation results for a hypothetical policy that has both an instantaneous and phase-in effect ($e^\beta_0=e^\beta_1=0.95$). The effect coded model is strongly biased towards the null and has low power, while the change coded model is has a small bias away from the null. Both the NB-DAM(1) and NB-DAM(2) are approximately unbiased. The NB-DAM(2) has higher power than its NB-DAM(1) counterpart (0.910 versus 0.807, respectively), and the lowest mean squared error of all approaches considered.  

\begin{figure}%
    \centering
    \includegraphics[width=\textwidth]{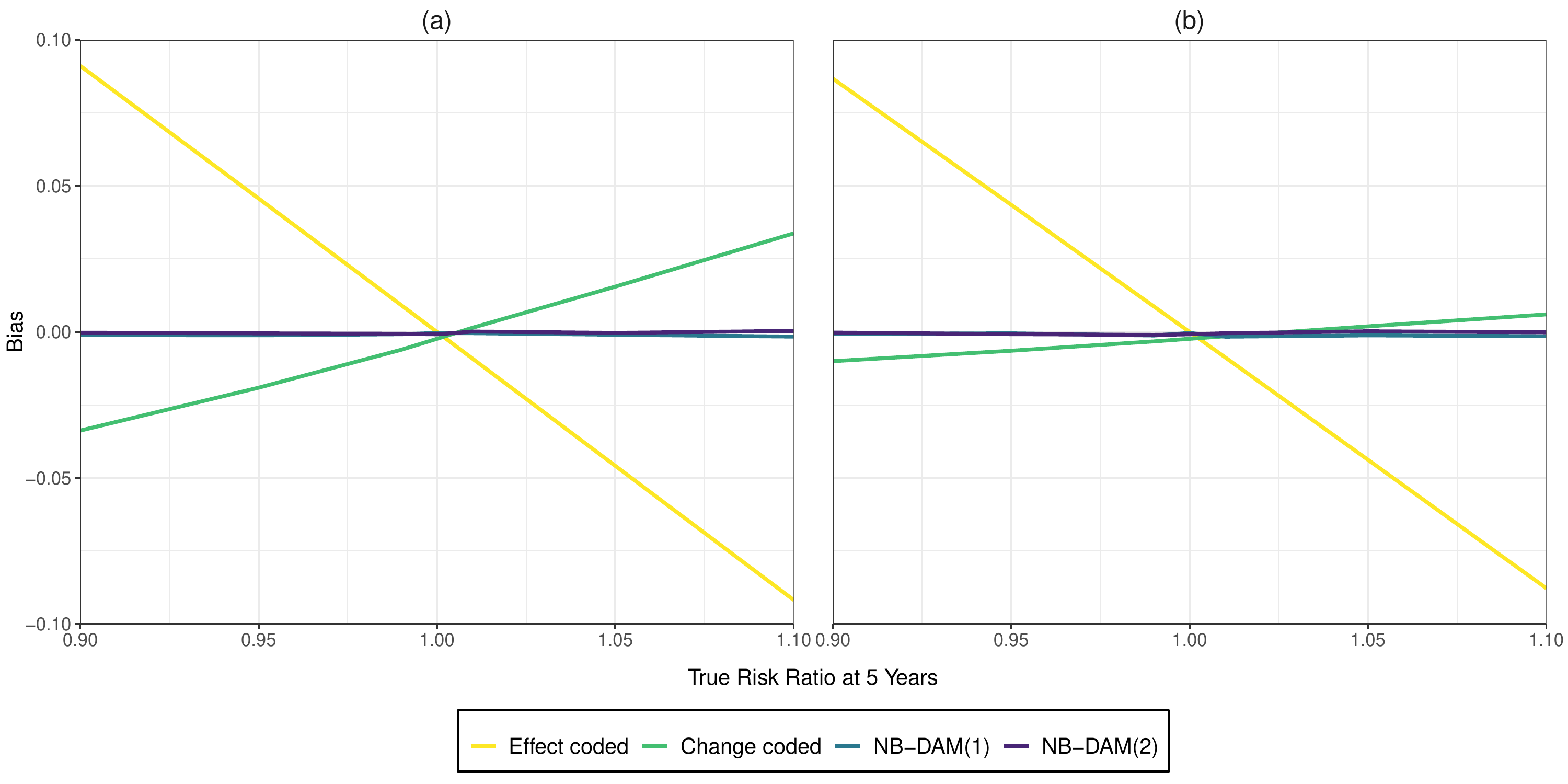} 
    \caption{Bias of different models from the simulation study. Panel (a) shows the bias at 5 years after implementation varying the true instantaneous effect when there is no phase in effect. Panel (b) shows the bias at 5 years after implementation varying the true phase in effect when there is no instantaneous effect.}
    \label{fig:bias}
\end{figure}

Figure \ref{fig:bias} illustrates each model's bias at different values of law effect parameters. Panel (a) show the bias 5-years after the law's implementation when varying the true instantaneous effect when there is no phase in effect ($e^{\beta_1}=1$). Both NB-DAMs are approximately unbiased, while the effect coded and change coded models are both biased. The effect coded model is biased towards the null, estimating a nearly null effect regardless of the true value. The change coded model is biased away from the null, with an increasing bias as the magnitude of the true effect increases. 

Panel (b) show the bias 5-years after the law's implementation when varying the true phase-in effect and imposing no instantaneous effect ($e^{\beta_0}=1$). Both NB-DAM models are approximately unbiased, while the effect coded and change coded models are both biased. As in Panel (a), the effect coded model is biased towards the null, estimating a nearly null effect regardless of the true value. The change coded model is slightly biased away from the null, with an increasing bias as the magnitude of the true effect increases. The biases shown in Figure \ref{fig:bias} (a) and (b) for the effect coded and change coded models are expected based on the strength of the autocorrelation (the unadjusted autocorrelation is estiamted at $0.82$). The more correlated the outcome is over time, the more bias towards the null the effect coded model will have, while the change coded model will become increasingly unbiased for stronger autocorrelation. 

\begin{table}[ht]
\centering
\begin{tabular}{|l|c|c|c|c|c|}
  \hline
Model & True RR & Bias & SD & MSE & Type I Error \\ 
  \hline
Effect coded & 1.00 & 0.0000 & 0.0072 & 0.0001 & 0.113 \\ 
  Change coded & 1.00 & -0.0024 & 0.0433 & 0.0019 & 0.030 \\ 
  NB-DAM(1) & 1.00 & -0.0009 & 0.0361 & 0.0013 & 0.070 \\ 
  NB-DAM(2)& 1.00 & -0.0012 & 0.0303 & 0.0009 & 0.062 \\ 
   \hline
\end{tabular}
\caption{Simulation results for estimating the effect of a policy at 5 years when there is no effect.}
\label{tab:null}
\end{table}

\begin{table}[ht]
\centering
\begin{tabular}{|l|c|c|c|c|c|}
  \hline
Model & True RR & Bias & SD & MSE & Power \\ 
  \hline
Effect coded & 0.9025 & 0.0867 & 0.0072 & 0.0076 & 0.416 \\ 
  Change coded & 0.9025 & -0.0215 & 0.0378 & 0.0019 & 0.769 \\ 
  NB-DAM(1) & 0.9025 & 0.0046 & 0.0329 & 0.0011 & 0.807 \\ 
  NB-DAM(2) & 0.9025 & 0.0034 & 0.0277 & 0.0008 & 0.910 \\ 
   \hline
\end{tabular}
\caption{Simulation results for estimating the effect of a policy at 5 years when there is both an instant and phase-in effect.}
\label{tab:power}
\end{table}

\begin{figure}%
    \centering
    \includegraphics[width=\textwidth]{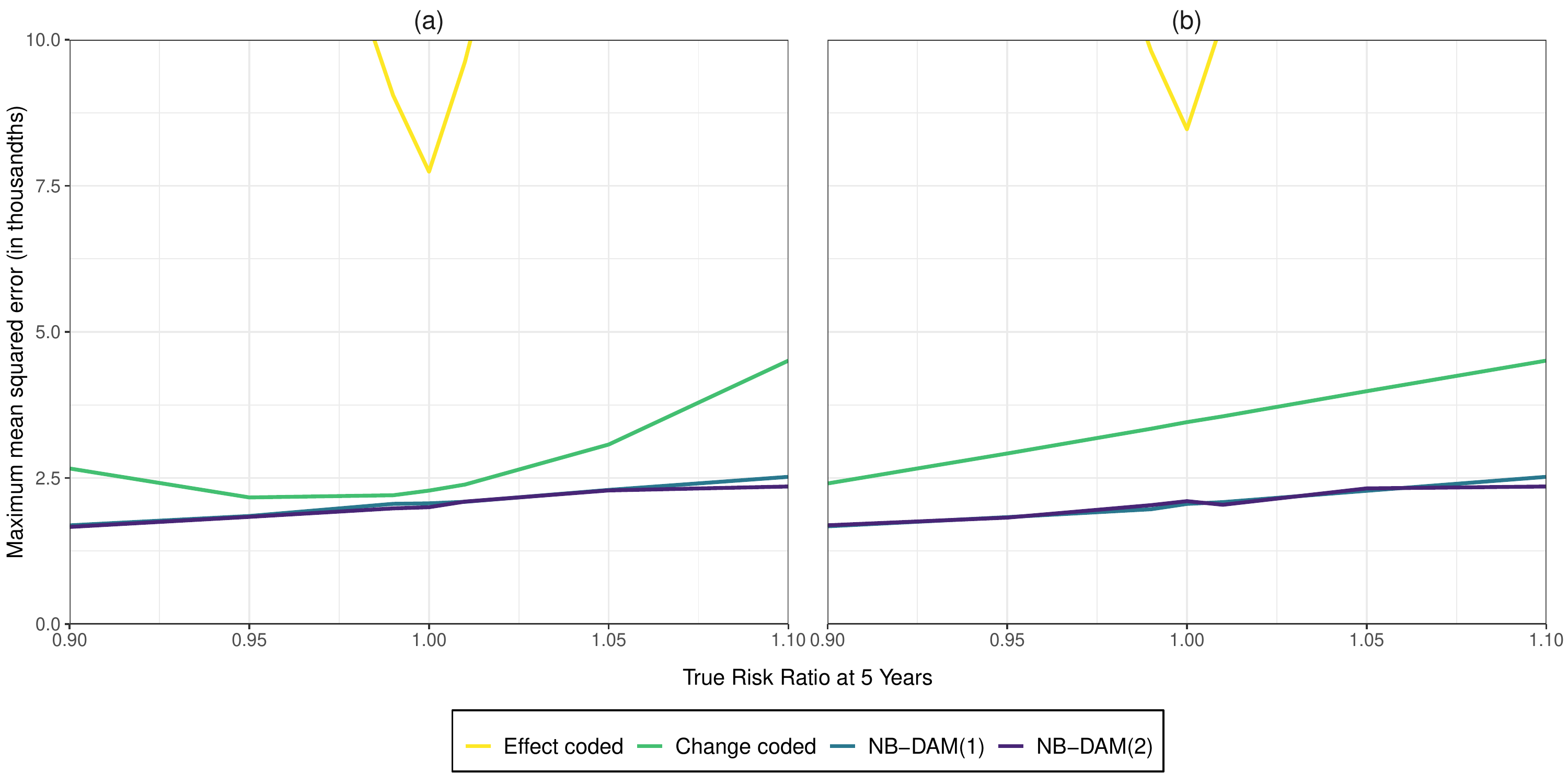}
    \caption{Mean squared error of different models from the simulation study. Panel (a) shows the maximum MSE across different true effects at 5 years after implementation with the true instantaneous effect on the x-axis. Panel (b) shows the maximum MSE across different true effects at 5 years after implementation with the true phase in effect on the x-axis. NOTE: The effect coded model is not shown as its MSE is out of the range of the figures.}
    \label{fig:mse}
\end{figure}

\section{Case Study: The Impact of Child Access Prevention Laws}

Using the same data described above for the simulation study, we estimate the effect of CAP laws on total firearm deaths. This is a simplification of the analysis described in \cite{schell2020} which  used these same data and estimated the effects of CAP and two other gun laws simultaneously. We code states as having CAP laws if they impose  civil or criminal penalties for storing a handgun in a manner that allows access by a minor according to the RAND State Firearm Law Database \citep{cherney2018development}. We consider two models. First, we estimate the negative binomial model of \cite{schell2020} that included four versions of the policy effect variables: the instant effect and five year phase-in effect described in (\ref{eqn:approx}), as well as their first-differences. Marginal effects were extracted from these models using numerical integration \citep{schell2020}. Second, we considered the NB-DAM(2) with $f_t(\cdot)$ defined by (\ref{eqn:approx}) with $b=5$. The risk ratio is estimated using the simple approximation in (\ref{eqn:est_effect}). Both models were fit using Stan in R \citep{rstan} with 10,000 MCMC iterations across four chains. 

\begin{figure}%
    \centering
    \includegraphics[width=.9\textwidth]{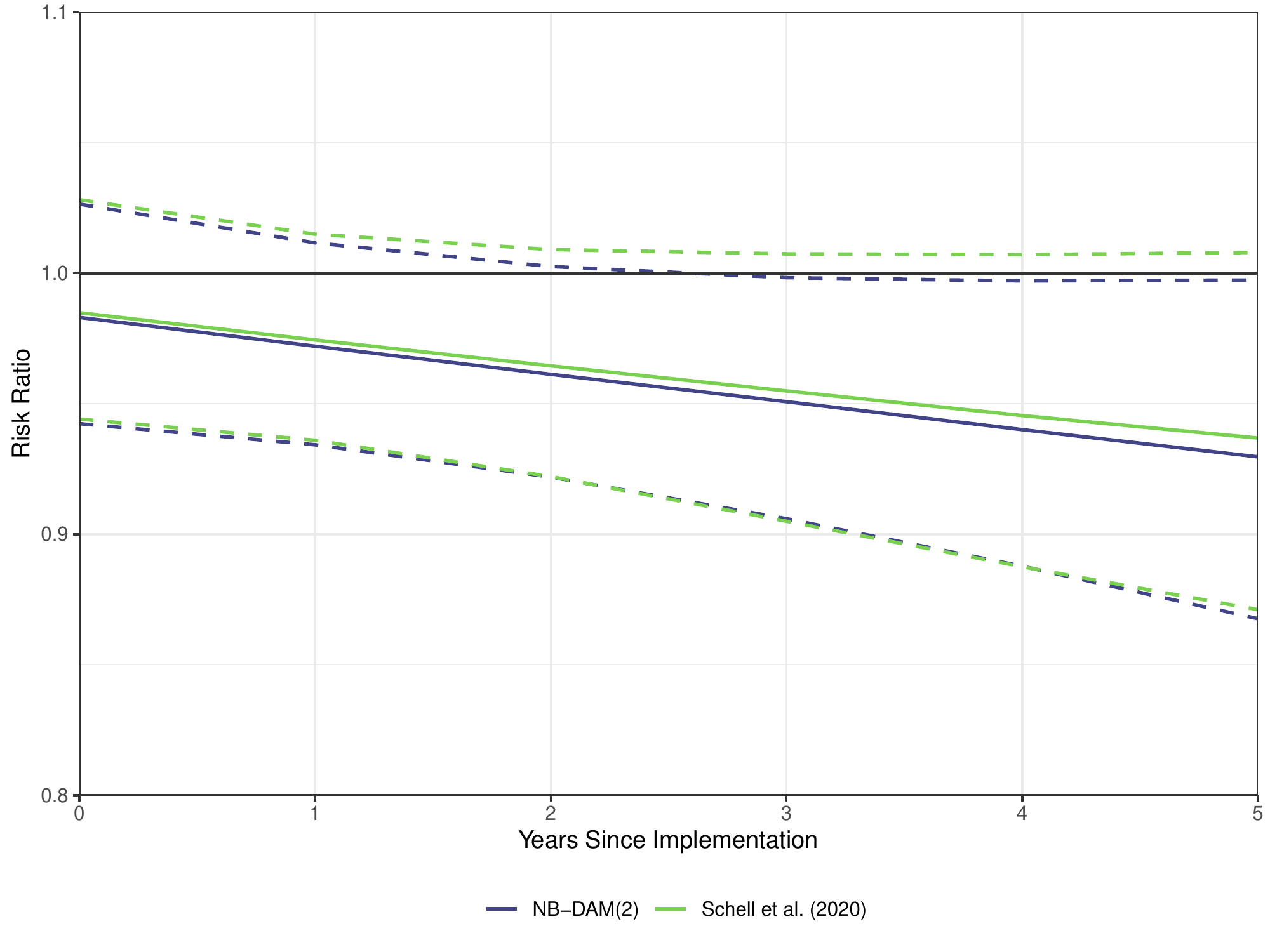} 
    \caption{Posterior median and 95\% credible intervals of the risk ratio measuring the impact of CAP laws on total firearm deaths. The posterior medians are represented by the solid lines, and the 95\% credible intervals are represented by the dashed lines.}
    \label{fig:analysis}
\end{figure}

Figure \ref{fig:analysis} plots the posterior median and 95\% credible intervals of the risk ratio measuring the impact of CAP laws on total firearm deaths. The posterior medians are represented by the solid lines, and the 95\% credible intervals are represented by the dashed lines. Both the posterior medians and the 95\% credible intervals are qualitatively similar, and the differences between the results from the two models is not substantively meaningful. However, the 95\% credible intervals for the risk ratio after 5 years from implementation for the NB-DAM(2) does not contain 1, 0.930 (0.868--0.997), while the interval for the model of \cite{schell2020} does, 0.937 (0.871--1.008). 

\section{Discussion}

This paper evaluates a class of autoregressive models for estimating effects in time series data that allow for direct interpretation of model parameters. The proposed approach is a novel application of these autoregressive models, rather than a post-estimation adjustment to the standard autoregressive model model. The parameters from a standard autoregressive model are not interpretable without additional computation because the model conditions on the lagged outcome, which may be endogenous to the variable of interest \citep{pearl2009causality}. Fortunately, for linear autoregressive models, the magnitude of this endogeneity can be computed at each time point as a function of model coefficients. The debiased autoregressive models of this paper effectively subtract the endogenous component from the models at each time point during estimation. 

The closed form results presented in this paper for linear autoregressive models are not generally available for other, nonlinear models. We propose a simple approximation that facilitates interpretation, and our empirical simulation study supports the use of such approximations when analyzing the effect of gun policies on total firearm deaths. We believe that total firearm deaths represents the expected behavior of many other outcomes when analyzing the effects of gun policy, but it is possible that the approximation is not suitable for outcomes in other domains. In light of that, we recommend that researchers who are concerned whether the proposed approach is appropriate for inference in their specific analysis conduct a simulation study within their dataset to verify the results of this study extend to their proposed analysis. Further research is required to better understand the limits of this approach for other applications.

\begin{appendix}
\section*{}
\subsection{Defining the Instant and Linear Phase-in Effects}

The outcomes discussed in this paper are measured as aggregations of events during specific windows in time, e.g., annual firearm deaths, and the state-level policies of interest may not align exactly with these measurement periods. As such, the effects of the policies may only be active during part of the year. To overcome this challenge, we parameterize our effects of interest using a continuous time scale, and then average these effects over the period of observation (on the log scale). Without loss of generality, assume that a policy becomes active at time $0 \leq t_e < 1$, where time is scaled to represent years. Assuming an instantaneous risk ratio of the policy in continuous time of size $exp(\beta_0)$, the average on the log risk ratio during the period between $t=0$ and $t=1$ is given by the fraction of year in which the policy was active, $(1-t_e)\beta_0$, and 1 for all time periods after $t=1$. 

For an effect that is assumed to phase-in linearly on the log scale, the form is only slightly more complex. Again, assume that a policy become active at time $0 \leq t_e < 1$, and assume a log risk ratio that phases-in linearly over a $b$ year period, i.e., $exp(\frac{t-t_e}{b}\beta_1)$. The effect at any time point is then given by:

\begin{align*}
    &1 &\text{     if } t<t_e \\
    &exp(\frac{t-t_e}{b}\beta_1) &\text{     if } t_e \leq t \leq b \\
    &exp(\beta_1) &\text{     if }  t > b 
\end{align*}

Averaging the log risk ratio between two subsequent years provides a closed form:

\begin{align*}
    &1 &\text{     if } t < 0 \\
    &exp( \frac{(1-t_e)^2}{2b}\beta_1) &\text{     if } t=0 \\
    &exp( \frac{t + 0.5 - t_e}{b} \beta_1) &\text{     if }  0 < t < b \\
    &exp( (1-\frac{t_e^2}{2b})\beta_1) &\text{     if } t=b \\
    &exp(\beta_1) &\text{     if } t>b \\
\end{align*}

\subsection{Additional Simulation Results}

\begin{figure}%
    \centering
    \includegraphics[width=.95\textwidth]{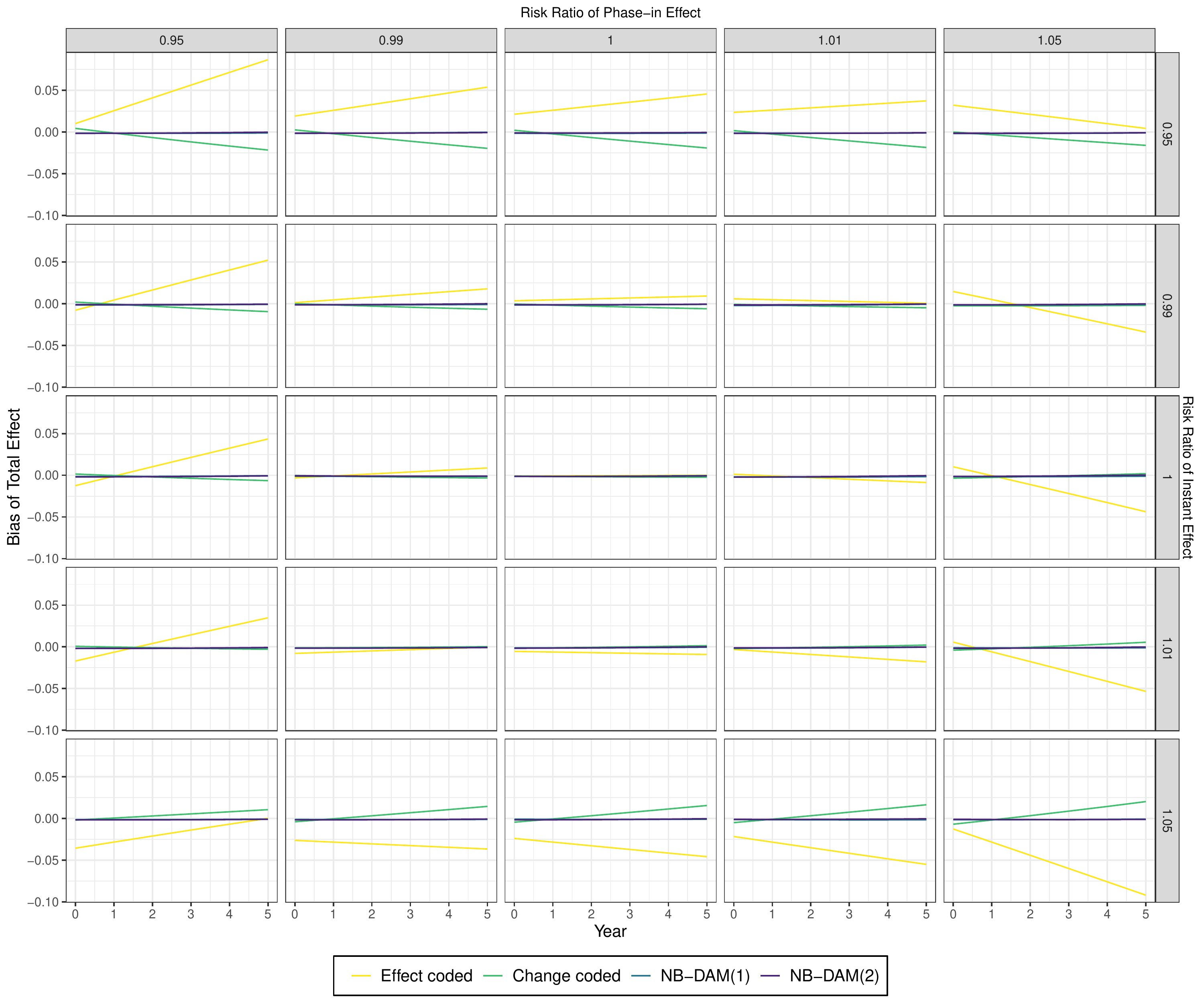}
    \caption{Bias for the total effect of a policy at different time points varying the true underlying instantaneous and phase-in effects.}
    \label{fig:bias-grid}
\end{figure}

Figure \ref{fig:bias-grid} provides the bias of different models for estimating the total effect of a policy at various time points after implementation. Of note, the NB-DAMs are approximately unbiased for all effects across all years, while the effect coded and changed coded models are not. 

\begin{figure}%
    \centering
    \includegraphics[width=.95\textwidth]{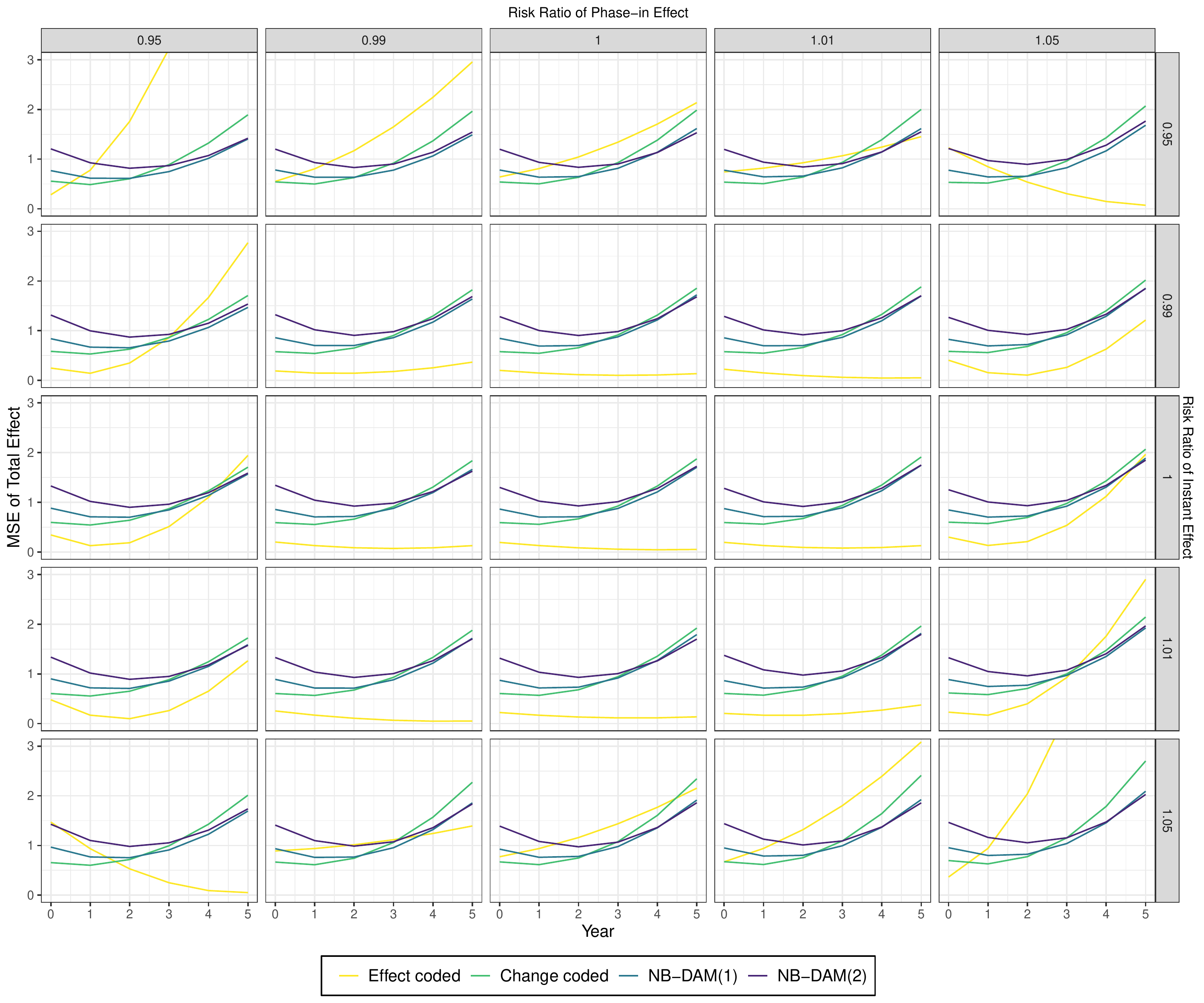}
    \caption{MSE for the total effect of a policy at different time points varying the true underlying instantaneous and phase-in effects.}
    \label{fig:mse-grid}
\end{figure}

Figure \ref{fig:mse-grid} provides the MSE of different models for estimating the total effect of a policy at various time points after implementation. The effect coded model has the lowest MSE when there is no true effect, but has worse performance as the size of the true effect increases. This occurs because the effect coded model is biased towards the null. The remaining models have similar patterns of MSE, with the change coded model tending to have lower MSE for estimating the instantaneous effect (those at Year 0), and the NB-DAMs having lower MSE for the total effect at 5 years after implementation.

\end{appendix}
\bibliographystyle{imsart-nameyear} 
\bibliography{biblio}       


\end{document}